# Weak field magnetoresistance of narrow-gap semiconductors InSb

R.Yang, G.L.Yu, Yanhui Zhang, P.P.Chen

*National Laboratory for Infrared Physics, Shanghai Institute of Technical Physics, Chinese Academy of Science, Shanghai 200083, People's Republic of China*

(abstract) The magnetoresistance of InSb has been intensively investigated. The experiments we perform here focus on weak field magnetoresistance of InSb thin film. We investigate the magnetoresistance of InSb films in perpendicular, tilted as well as parallel magnetic field. Our results verify the previous observations concerning weak localization effect in InSb thin film. Moreover, we systematically study the anisotropy of magnetoresistance of InSb. We find that the existence of in-plane field can effectively suppress the weak localization effect of InSb film. We fit the experimental data with two types of models, the match between data and model is excellent. From the fitting procedure, we get information about phase coherence time, spin-orbit scattering time. The information about Zeeman effect and sample roughness are also extracted from the fitting procedure.

Ⅰ.Introduction

Weak localization effect happens when $L_\phi > L_r$, $L_\phi$ is phase coherence length and $L_r$ is the average length that a carrier travels before returning to its original position. When the crystal is perfect and there is no disorder, it's basically impossible for a carrier to return to its original position, thus $L_r$ tends to be infinity and weak localization effect can't happen. When there is moderated amount of disorder, so carrier can return to its original position through scattering and the inelastic collision has not drastically reduced $L_\phi$, thus weak localization can happen in this case. The further increase of disorder will reduce weak localization effect, because too much disorder will cause too many inelastic collisions, thus drastically reduce $L_\phi$, and weak localization will vanish as soon as the condition $L_\phi > L_r$ breaks. According to the above analysis, it's obvious that the strength of weak localization will firstly increase, then decrease with the increase of disorder.

The weak localization effect with strong spin-orbit interaction in InSb film has been observed in many experiments[1-4]. The phase coherence time and spin-orbit scattering time are obtained from the analysis of weak localization effect observed in our InSb samples. Moreover, though the anisotropy of weak localization effect with respect to the direction of applied magnetic field has been observed[2,4], to our best knowledge, there are no researches dedicated to this anisotropy up to date. Here, we systematically study the anisotropy of weak field magnetoresistance of intrinsic InSb film and explain it under two possible scenarios. There are many scenarios concerning the origin of the weak localization effect. The models of weak localization effect fall into three types. One type is weak localization models of 2-D system[5,6]. Another type is weak localization models of quasi-2D system which differs from models of 2-D system by replacing diffusion coefficient、carrier density and resistivity with corresponding 3D value. The third model is weak localization models for 3D system[7,8]. Here, we analyze the data with quasi-2D and 3D models.

Ⅱ.Sample fabrication and measuring system

The InSb film used here were grown on semi-insulating GaAs <100> substrate by MBE facility. The thickness of InSb film is 1 micrometer. In is deposited onto InSb surface to facilitate Ohmic contacts. The magnetoresistance is measured under Van der Pauw configuration and the magnetic field is applied perpendicular, tilted and parallel to the film. All measurements are performed in an Oxford Instruments cryogenic system with temperature ranges from 1.3K~2.1K.

Ⅲ. Experimental results and discussion

In perpendicular magnetic field:

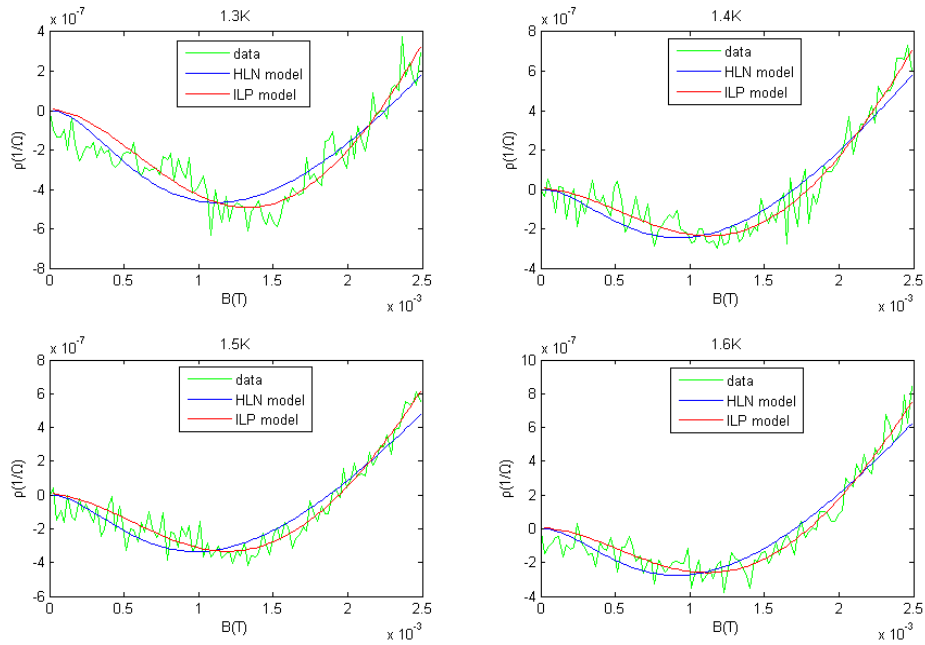

a

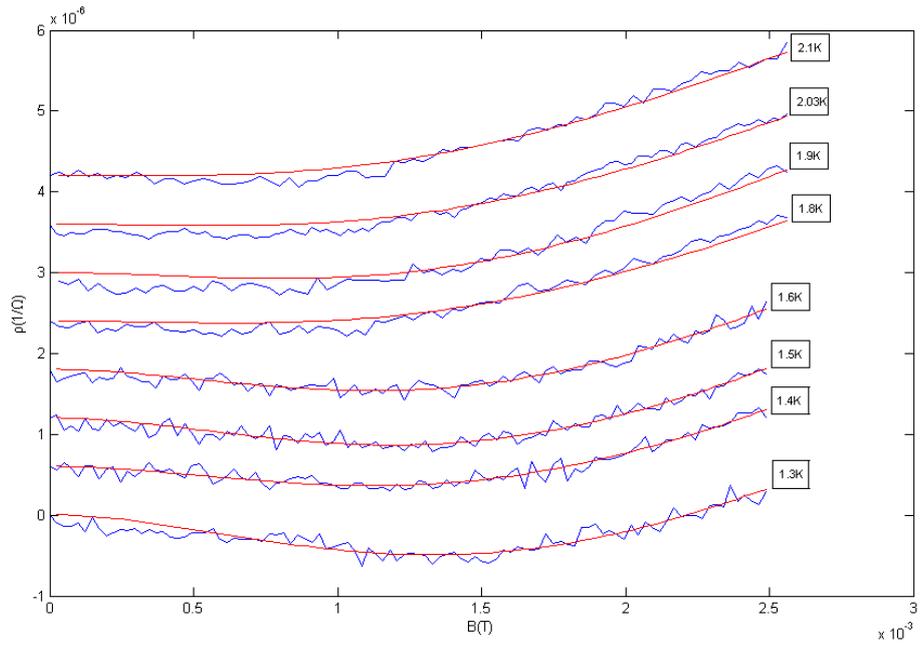

b

FIG.1: (a) Fitting results with 2D models. Green curves are experimental data at varies temperature; blue curves are fitting with HLN model; red curves are fitting with ILP model; (b)Comparison of fitting results corresponding to different temperatures, all the curves are vertically shifted for clarity.

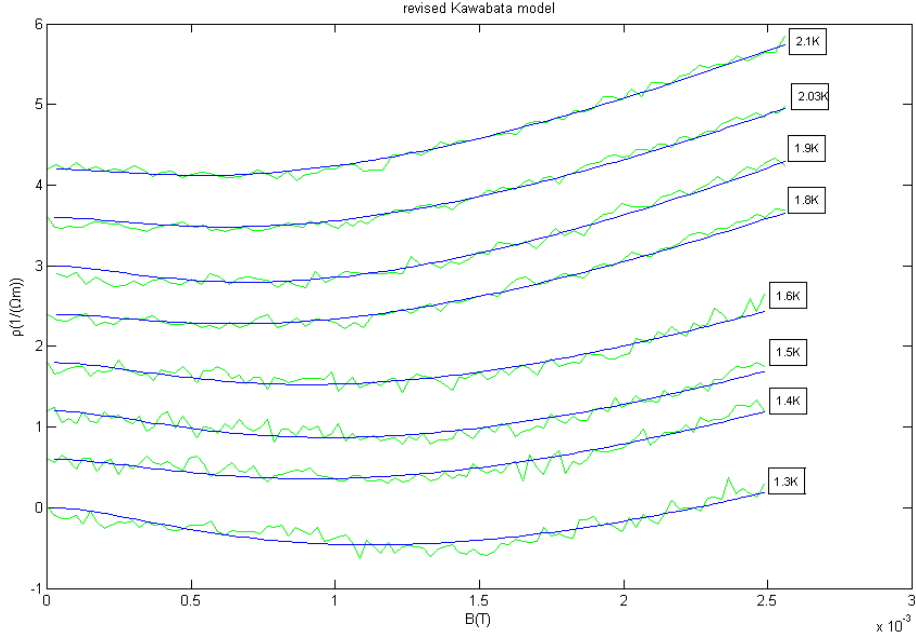

FIG2: Fitting results with 3D model (revised Kawabata model). Green curves are experimental data at varies temperature; blue curves are fitting with revised Kawabata model.

We can see clearly the weak localization effect with strong spin-orbit effect from these pictures. The negative magnetoconductivity before the emerging of positive magnetoconductivity is a mark of strong spin-orbit interaction. Moreover, with the increase of temperature, we can see that the negative magnetoconductivity 'cusp' fades away gradually. This is because the depth of the 'cusp' is proportional to $\tau_\varphi / \tau_{so}$, with the increase of temperature, $\tau_\varphi$ decreases and $\tau_{so}$ is temperature-independent. This phenomenon is another feature of weak localization effect with spin-orbit effect.

There are two types of models that are applicable here; one type is weak localization model in 2D system and which is used by many research groups when it comes to the weak localization in narrow-band semiconductors[1,5,6]. The other type is weak localization model in 3D system[2,3,7,8,11], here, we use revised Kawabata's model[2].

The weak localization model in 2D system has varies versions. Here, we use two widely-used versions (FIG.1). In addition, there is some controversy about which model should be used in InSb system. When it comes to the weak localization effect in InSb thin film, some groups regard thin InSb film as a quasi-2D system and apply the 2D models[1]; on the other hand, other groups regard it as a 3D system and apply 3D models[2,3,11]. In the following part, we will also compare these different approaches in detail.

In 2D system, the models concerning weak localization effect with strong spin-orbit effect have a general form: $\Delta\sigma(B) - \Delta\sigma(0) = f(B; \tau_\varphi, \tau_{so})$

$\tau_\varphi, \tau_{so}$ are phase coherence time and spin-orbit scattering time separately.

In 2D regime, we fit the experimental data with two models—revised HLN model[1] which takes spin-orbit interaction into consideration and ILP model[6,10].

For the revised HLN model[1],

$$\Delta\sigma(B) = -\frac{e^2}{2\pi^2\hbar}\left[\psi\left(\frac{1}{2} + \frac{B_{tr}}{B}\right) + \frac{1}{2}\psi\left(\frac{1}{2} + \frac{B_\varphi}{B}\right) - \frac{3}{2}\psi\left(\frac{1}{2} + \frac{B_\varphi}{B} + \frac{4}{3}\frac{B_{so}}{B}\right)\right]$$

$\psi(x)$ is the digamma function; $B_{tr} = \frac{\hbar}{4eD\tau_p}$; $B_\varphi = \frac{\hbar}{4eD\tau_\varphi}$; $B_{so} = \frac{\hbar}{4eD\tau_{so}}$

It only considers E-Y spin dephasing mechanism[6]. The fitting results according to this model are the blue curves in FIG.1.

For ILP model[10],

$$\Delta\sigma(B) - \Delta\sigma(0) = \frac{G_0}{2}\left(F_t(b_\varphi, b_s) - F_s(b_\varphi)\right),$$

$$F_s(b_\varphi) = \psi(\frac{1}{2} + b_\varphi) - \ln b_\varphi$$

$$F_t(b_\varphi, b_s) = \sum_{n=1}^{\infty}\left\{\frac{3}{n} - \frac{3a_n^2 + 2a_n b_s - 1 - 2(2n+1)b_s}{(a_n + b_s)a_{n-1}a_{n+1} - 2b_s\left[(2n+1)a_n - 1\right]}\right\} - \frac{1}{a_0} - \frac{2a_0 + 1 + b_s}{a_1(a_0 + b_s) - 2b_s}$$
$$-2\ln(b_\varphi + b_s) - \ln(b_\varphi + 2b_s) - 3C - S(b_\varphi/b_s)$$

$$G_0 = \frac{e^2}{2\pi^2\hbar}; b_\varphi = \frac{B_\varphi}{B}; b_s = \frac{B_s}{B}; S(x) = \frac{8}{\sqrt{7+16x}}\left[\arctan(\frac{\sqrt{7+16x}}{1-2x}) - \pi\Theta(1-2x)\right]$$

$$a_n = n + \frac{1}{2} + b_\varphi + b_s;$$

$$B_\varphi = \frac{\hbar}{4eD\tau_\varphi}; B_s = \frac{\hbar}{4eD\tau_{so}}$$

$\psi(x)$ is the digamma function

C is the Euler constant, $\Theta(x)$ is the Heaviside step function

It considers both E-Y and D-P spin dephasing mechanisms[6]. In the fitting procedure, we omit the $\Omega_3$ term, because in our film sample, the Dresshaul term should dominate the spin-orbit interaction[9][10]. The fitting results according to this model are the red curves in FIG.1.

As can be seen from FIG.1, fitting according to ILP model is better. That means both EY and DP spin dephasing mechanisms are important in InSb thin film considered here.

In 3D regime, we apply revised Kawabata model[2] which takes into consideration the spin-orbit coupling.

$$\Delta\sigma(B) - \Delta\sigma(0) = \left(\Delta\sigma_s(B) - \Delta\sigma_s(0)\right) + \left(\Delta\sigma_t(B) - \Delta\sigma_t(0)\right),$$

$$\Delta\sigma_s(B) - \Delta\sigma_s(0) = \frac{e^2}{4\pi^2\hbar l_B} \sum_{n=0}^{\infty} \left[ P_{1/2}^{-1} \cos(\frac{\varphi_1}{2}) - 2P_1 \cos(\frac{\varphi_2}{2}) + 2P_0 \cos(\frac{\varphi_0}{2}) \right]$$

$$\Delta\sigma_t(B) - \Delta\sigma_t(0) = -\frac{3e^2}{4\pi^2\hbar l_B} \sum_{n=0}^{\infty} \left[ (n + \frac{1}{2} + \delta_{so})^{-\frac{1}{2}} - 2(\sqrt{n+1+\delta_{so}} - \sqrt{n+\delta_{so}}) \right]$$

$$P_s = \left[ (n+s+\delta)^2 + \upsilon^2 \right]^{\frac{1}{4}}, \delta = \frac{l_B^2}{4D\tau_\varphi}, \delta_{so} = \frac{l_B^2}{4D}\left(\frac{1}{\tau_\varphi} + \frac{4}{3\tau_{so}}\right)$$

$$\upsilon = \frac{g\mu_B}{4eD}, \varphi_s = \arctan\left[\frac{\upsilon}{n+s+\delta}\right]$$

$$l_B = \sqrt{\frac{\hbar}{eB}}$$

We can see the fitting with revised Kawabata model is also excellent (FIG.2).

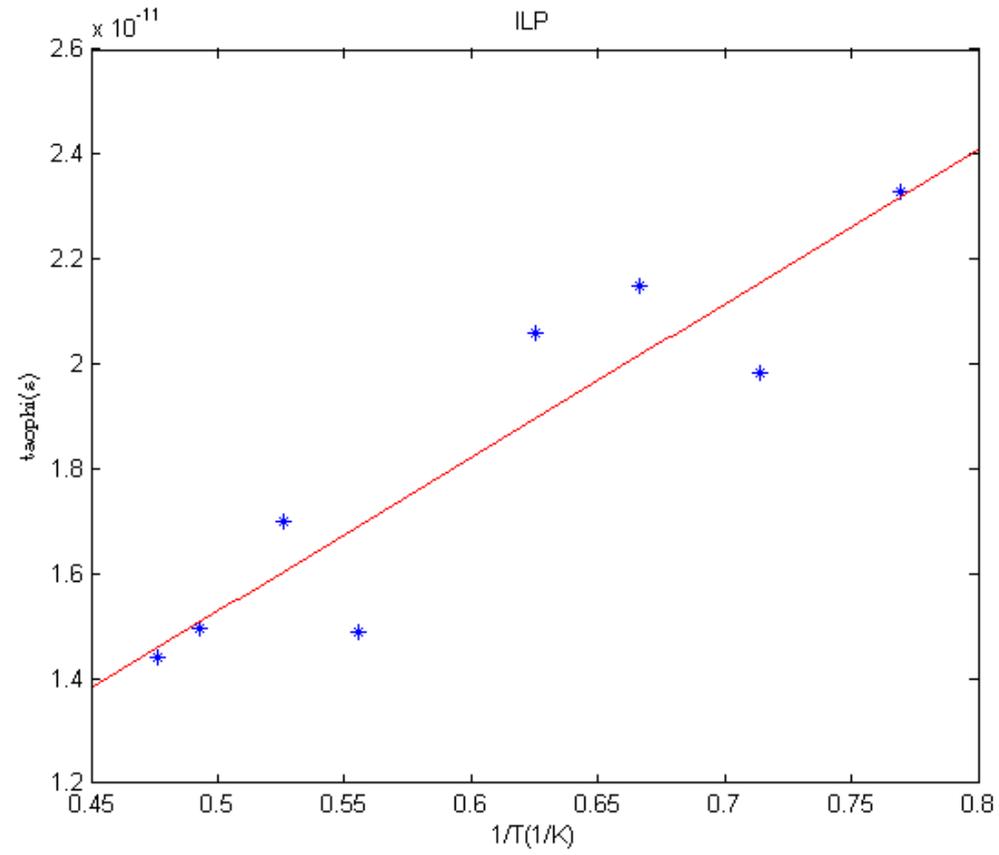

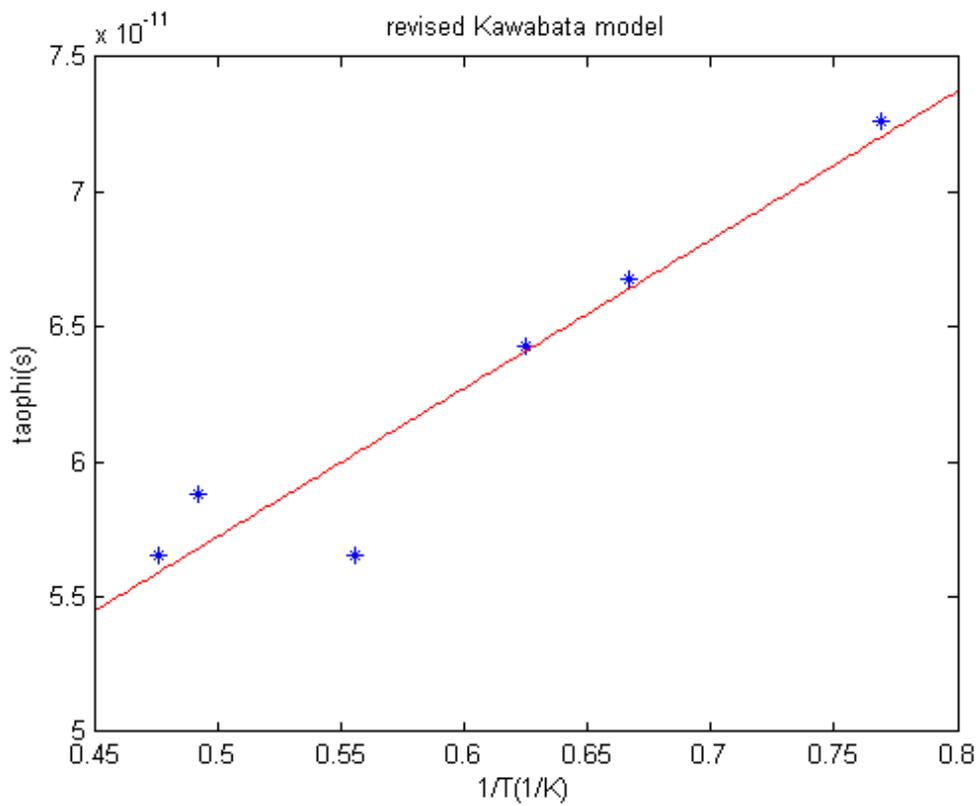

FIG.3: plots of $\tau_\varphi$ vs $1/T$, in the upper picture $\tau_\varphi$ is got from fitting data with ILP

model; in the lower picture, $\tau_\varphi$ is got from fitting data with Kawabata's model.

From the relation between $\tau_\varphi$ and T, we can get the information about the decoherence mechanism in InSb thin film. Generally speaking, without saturation effect, $\tau_\varphi \sim \left(\frac{1}{T}\right)^p$, and p=1 and p=3 are corresponding to Nyquist decoherence mechanism[1] and electron-phonon interaction decoherence mechanism separately[12,13]. For our data, the best fit is obtained for p=1, corresponding to Nyquist decoherence mechanism. The result is consistent with previous research[1].

The Nyquist decoherence mechanism[1] predicts a linear relation between $\tau_\varphi$ and $\frac{1}{T}$:

$$\tau_\varphi = \alpha \cdot \frac{1}{T}, \alpha = \frac{\frac{2\pi\hbar^2}{e^2}\sigma_0 t}{k_B \ln\left(\frac{\pi\hbar}{e^2}\sigma_0 t\right)}$$

With $\tau_\varphi$ got from ILP model, after fitting $\tau_\varphi \sim \frac{1}{T}$ curve with linear equation, we can see that data corresponding to ILP model give rise to decoherence rate (0.3e-10) close to theory value(1.87e-10). With the $\tau_\varphi$ got from revised Kawabata model, we can get decoherence rate (0.6e-10) closer to theoretical value.

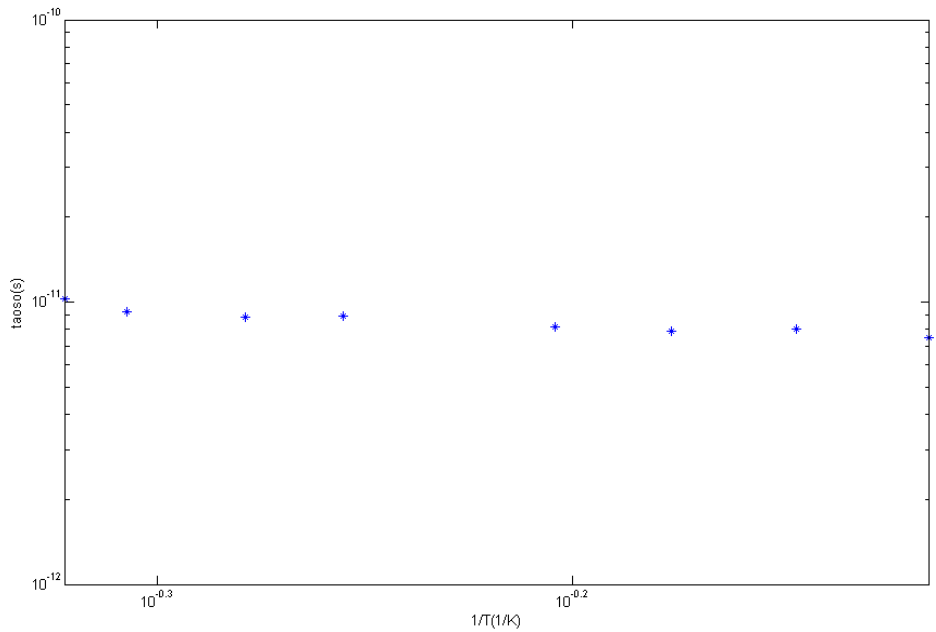

$\tau_{so}$ is basically temperature-independent, both EY and DP mechanisms can lead to temperature-independent $\tau_{so}$, so, in order to distinguish between these two mechanisms, it's necessary to analysis the dependence of $\tau_{so}$ on n and $\tau_p$ [1].

Both 2D model (ILP model) and 3D model (revised Kawabata's model) can give rise to good fitting and acceptable decoherence rate in above analysis; to decide which one is better requires further investigation. In the remaining part, the anisotropy of weak localization effect observed in our experiment will be investigated.

In tilted magnetic field

The anisotropy of weak localization effect has been predicted[15] and observed[10,14] in quantum wells. The anisotropy has been systematically investigated in InGaAs and GaAlN quantum wells[10,14,16]. To our best knowledge, it has not been investigated in detail in thin film system.  Here, we investigate this phenomenon in detail.

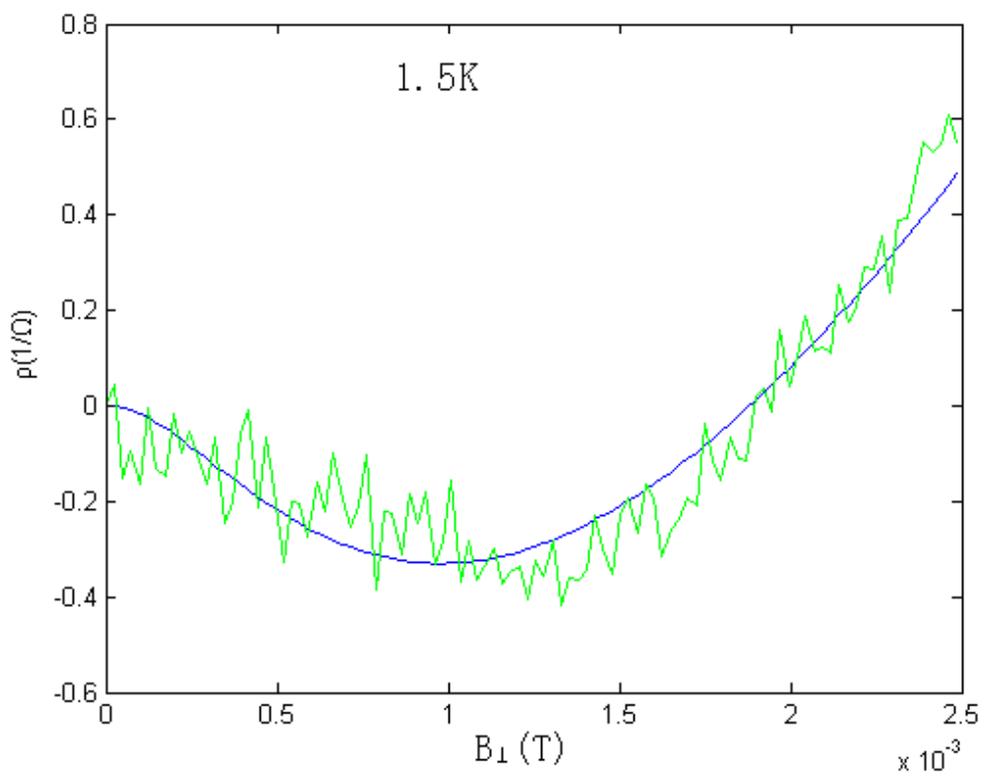

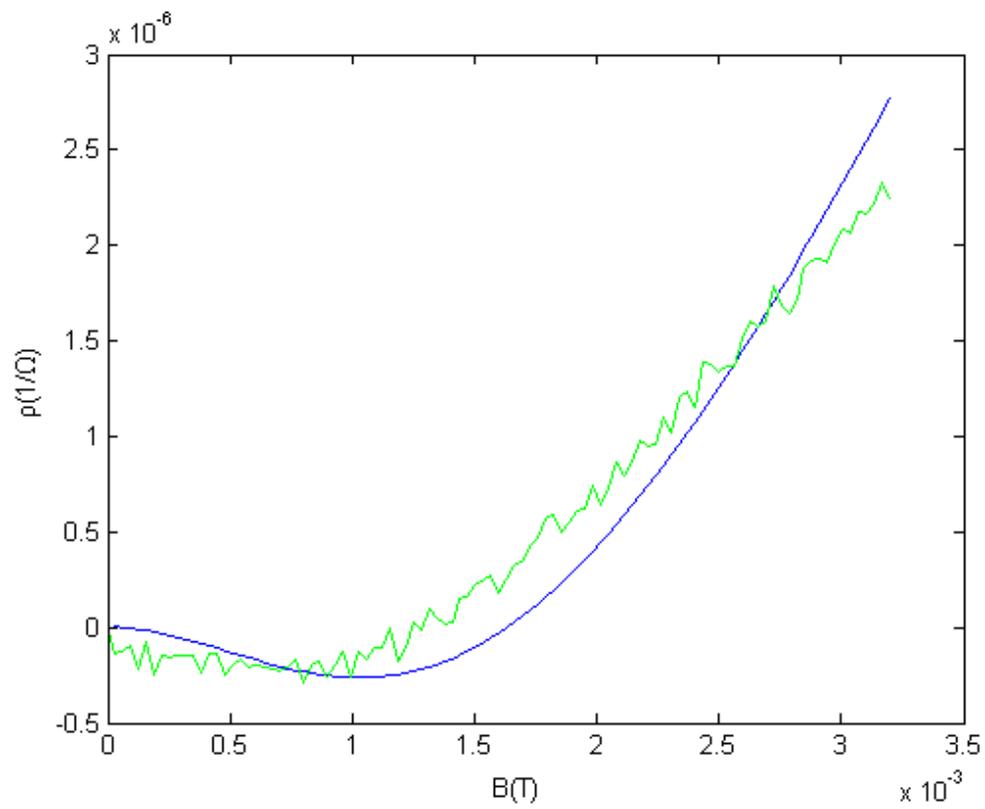

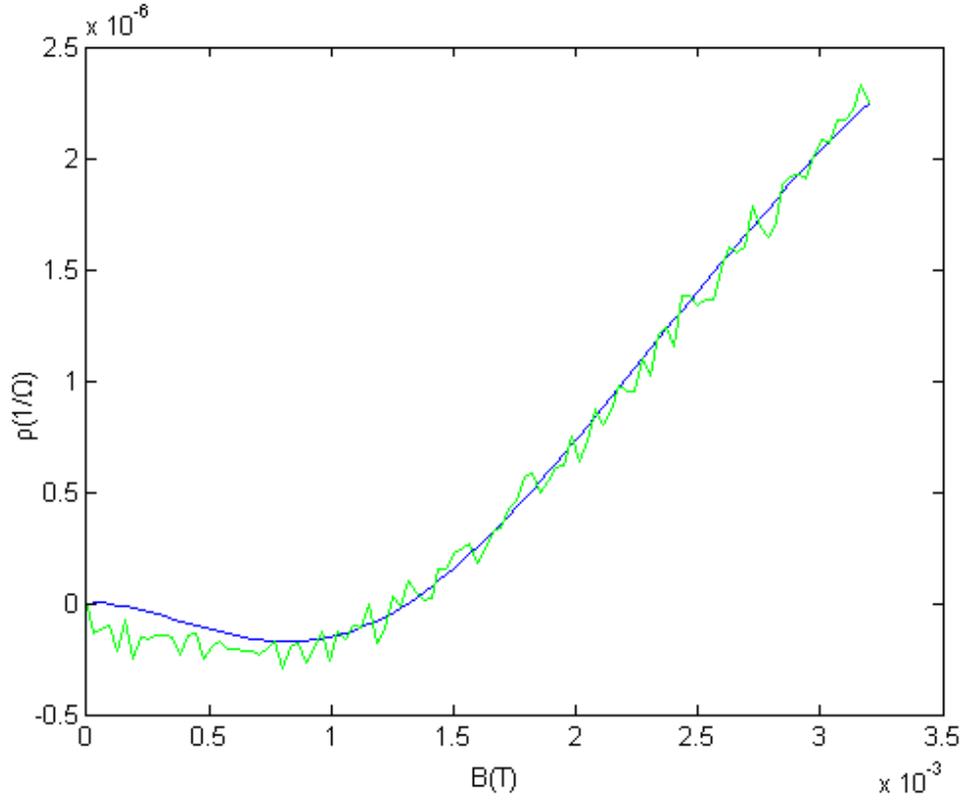

FIG.4: The up plot corresponds to fitting with ILP model under tilted magnetic field; The middle and bottom plots correspond to tilted magnetic field configuration with theta=30 degree. The blue curve in middle plot is fitting with model that only takes Zeeman interaction into consideration. The blue curve in the bottom plot is fitting with model that takes both Zeeman and roughness effect into consideration.

We can see from FIG.4 that the weak antilocalization effect (the cusp-like negative magnetoconductivity part of the magnetoconductivity curve) is suppressed in tilted magnetic field. This is an effect that has been widely observed in previous magnetoresistance experiment of quantum well systems[10,14,16].

There are two scenarios can be used to explain our experiment data.

The first scenario is the quasi-2D model:

The basic physics behind this phenomenon is that the in-plane field can lead to the suppression of weak antilocalization effect through Zeeman effect and roughness effect[10 16]. We fit our data with models proposed in reference 10 that investigates the suppression of weak antilocalization of InGaAs quantum well in a tilted magnetic field.

According to reference 10, the magnetoconductivity of 2DEG in a tilted field can be expressed as:

$$\Delta\sigma(B) - \Delta\sigma(0) = f(B_\perp; g, d^2 L, \tau_\varphi, \tau_{so}) = \frac{G_0}{2}\left(F_t(b_\varphi, b_s) - F_s(\widetilde{b_\varphi})\right),$$

$$F_s(\widetilde{b_\varphi}) = \psi(\frac{1}{2} + \widetilde{b_\varphi}) - \ln\widetilde{b_\varphi}$$

$$F_t(b_\varphi, b_s) = \sum_{n=1}^{\infty}\left\{\frac{3}{n} - \frac{3a_n^2 + 2a_n b_s - 1 - 2(2n+1)b_s}{(a_n + b_s)a_{n-1}a_{n+1} - 2b_s\left[(2n+1)a_n - 1\right]}\right\} - \frac{1}{a_0} - \frac{2a_0 + 1 + b_s}{a_1(a_0 + b_s) - 2b_s}$$
$$-2\ln(b_\varphi + b_s) - \ln(b_\varphi + 2b_s) - 3C - S(b_\varphi/b_s)$$

$$G_0 = \frac{e^2}{2\pi^2\hbar}; b_\varphi = \frac{B_\varphi + \Delta_r(B_\parallel)}{B_\perp}; b_s = \frac{B_s}{B}; \widetilde{b_\varphi} = \frac{B_\varphi + \Delta_r(B_\parallel) + \Delta_s(B_\parallel)}{B_\perp};$$

$$S(x) = \frac{8}{\sqrt{7 + 16x}}\left[\arctan(\frac{\sqrt{7 + 16x}}{1 - 2x}) - \pi\Theta(1 - 2x)\right]$$

$$a_n = n + \frac{1}{2} + b_\varphi + b_s;$$

$$\Delta_r(B_\parallel) \simeq \frac{\sqrt{\pi}}{2}\frac{e}{\hbar}\frac{d^2 L}{l}B_\parallel^2; \Delta_s(B_\parallel) = \frac{\tau_s}{4e\hbar D}\left(g\mu_B B_\parallel\right)^2;$$

$$B_\varphi = \frac{\hbar}{4eD\tau_\varphi}; B_s = \frac{\hbar}{4eD\tau_{so}}$$

$\psi(x)$ is the digamma function

C is the Euler constant, $\Theta(x)$ is the Heaviside step function

g - the Lande factor; $d^2L$ - a value that measures the roughness.

Generally, fitting data with this formula is a surface fitting problem. In our experiment, $B_\parallel = B_\perp \tan\theta$, so it can be simplified into a curve fitting problem.

$$\Delta\sigma(B_\perp, B_\parallel) - \Delta\sigma(0, B_\parallel) = f'(B_\perp; g, d^2L, \tau_\varphi, \tau_{so})$$

In FIG.4, we can see the fitting is quite well. The fitting result of the middle plot is:

g=323

The fitting results of the bottom plot are: g=500; $d^2L$=5.8367e-21(the fluctuation volume is ~(100nm)^3)

So, in the framework of 2D model, both Zeeman Effect and roughness effect are important when it comes to the weak field magnetoresistance anisotropy of InSb film.

The problem confronted here is that the g factor got from fitting is 10 times larger than the intrinsic g factor (~50) of InSb. A possible explanation is that due to $L_\phi$~0.5um is comparable with film thickness, such that the effect of in-plane field on electron's orbit motion cannot be neglected ($r_{orbit}$~0.1nm).

The second scenario is the 3D model (revised Kawabata model):

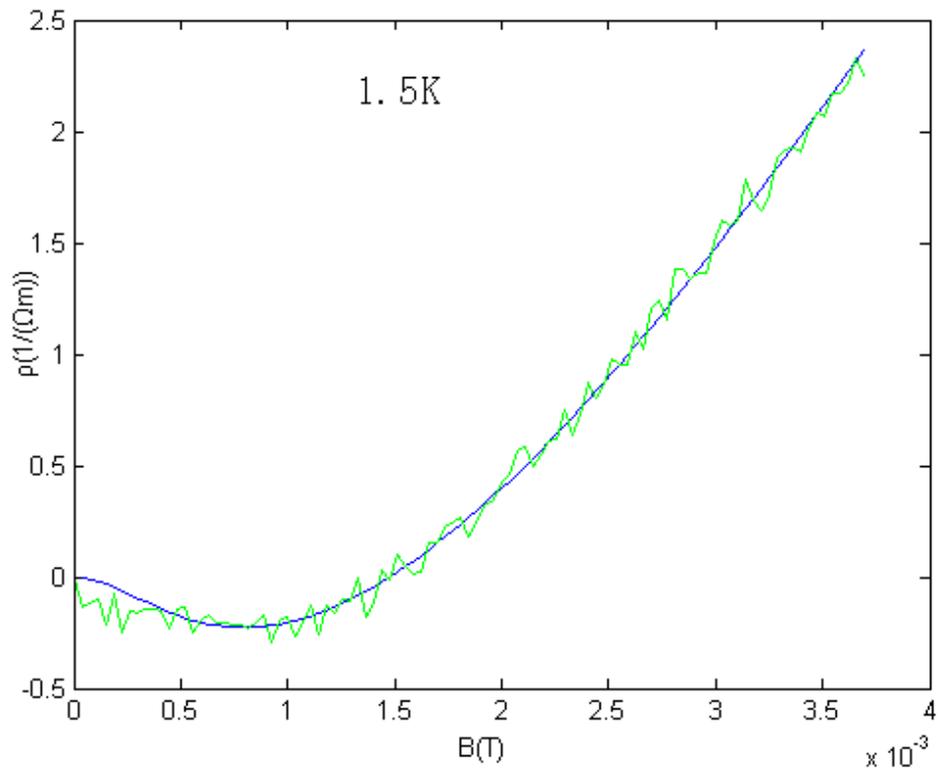

FIG.5

From FIG.5, we can see that the fitting with 3D model is even more excellent. Another advantage of this model it that the g factor used here is 51, the same as intrinsic g factor of InSb. However, the anisotropy observed here remains a mystery. A possible explanation with 3D model is that, in contrast to common sense, $\tau_\varphi$ and $\tau_{so}$ in a bulk material are intrinsically anisotropic. Another problem met here is that the validity of 3D model, the 3D model is valid for $L\varphi \ll t$, t is the thickness, however, in our sample, $L\varphi$=0.5um, t=1um, $L\varphi$~1/2t.

In parallel magnetic field

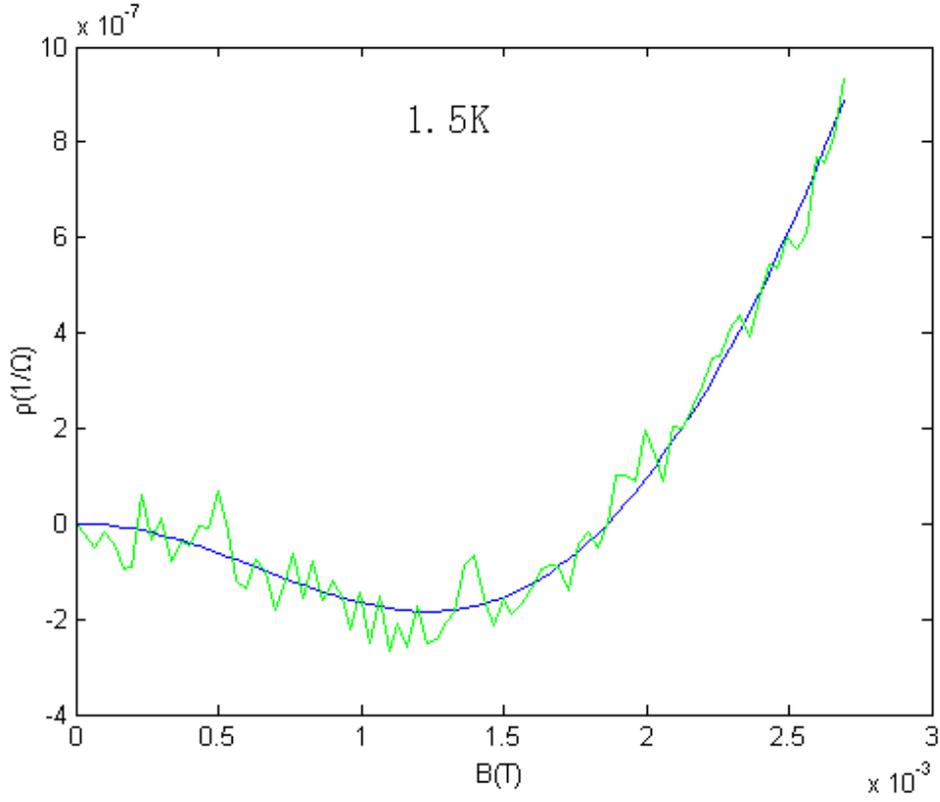

FIG.6

From FIG.6, we can see that the small positive magnetoresistance 'cusp' becomes obvious again in parallel magnetic field. But it's still weaker than that in perpendicular magnetic field.

In 2D scenario, fit the experimental data with model in reference 10 with the configuration of parallel magnetic field:

$$\Delta\sigma(B)-\Delta\sigma(0)=\frac{G_0}{2}\left(2\ln(\frac{B_\varphi+B_s+\Delta_r}{B_\varphi+B_s})+\ln(\frac{B_\varphi+2B_s+\Delta_r}{B_\varphi+2B_s})-\ln(\frac{B_\varphi+\Delta_s+\Delta_r}{B_\varphi})+S(B_\varphi/B_s+\Delta_r/B_s)-S(B_\varphi/B_s)\right),$$

$$\Delta_r(B_\parallel) \simeq \frac{\sqrt{\pi}}{2} \frac{e}{\hbar} \frac{d^2 L}{l} B_\parallel^2; \Delta_s(B_\parallel) = \frac{\tau_s}{4e\hbar D}(g\mu_B B_\parallel)^2;$$

$$B_\varphi = \frac{\hbar}{4eD\tau_\varphi}; B_s = \frac{\hbar}{4eD\tau_{so}}$$

We can get parameters very close to previous values.

g=500, $d^2 L$ =3.8155e-21, $\tau_\varphi$ =1.1306e-11, $\tau_{so}$ =7.0551e-12

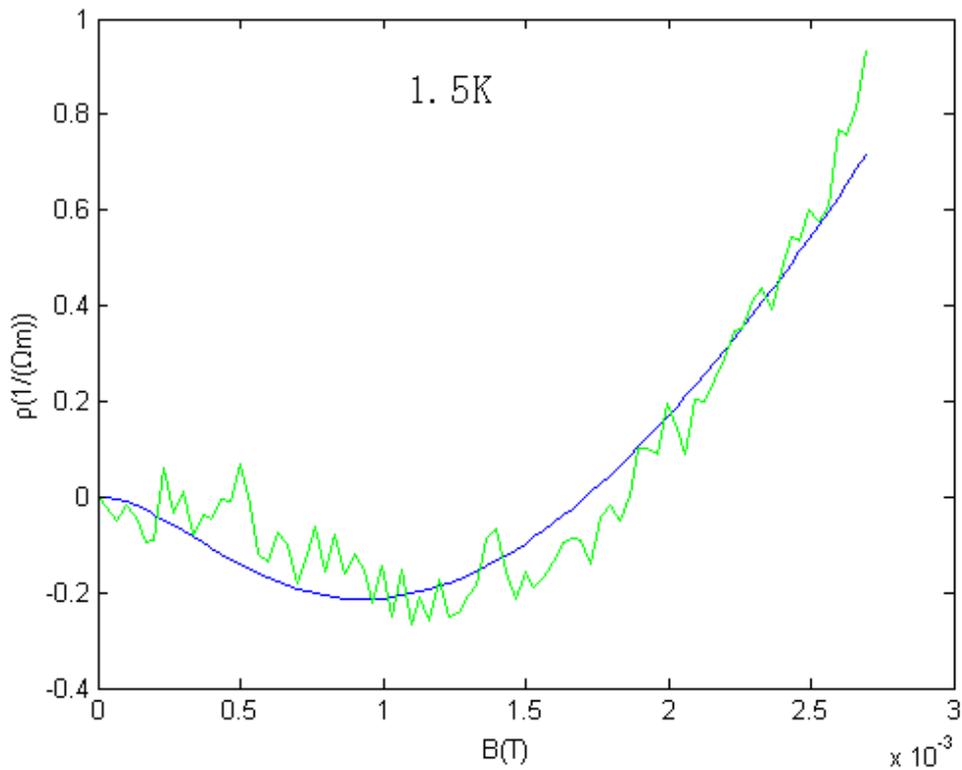

FIG.7: fitting result of the data corresponding to parallel magnetic field configuration with revised Kawabata's model

In 3D scenario, the fitting result with revised Kawabata's model is shown in FIG.7.

And from fitting procedure, we get that $\tau_\varphi$=5.1803e-11s, $\tau_{so}$=2.5188e-11s. These values are consistent with the values got from perpendicular-field configuration. We can see that the fitting curve is not as good as that of 2D scenario. This also implicates that the application of 3D scenario is not appropriate here.

Ⅳ.Conclusions

We investigate the weak localization effect in InSb thin film. We compare different models concerning weak localization effect in detail and illustrate the appropriateness of quasi-2D model used in this research. By fitting the data with model, we also get the conclusion that Nyquist mechanism is responsible for decoherence in our sample. We also systematically study the anisotropy of weak localization effect with respect to magnetic field. We find that the existence of in-plane field can effectively suppress the weak localization effect of InSb film. Moreover, the anisotropy has been explained with 3D and 2D models, and the implications and appropriateness of different models are also discussed.

**References**

1 R. L. Kallaher, J. J. Heremans, Phy. Rev. B **79**, 075322 (2009)

2 M. Oszwaldowski, T. Berus, Phy. Rev. B **65**, 235418 (2002)

3 R. C. Dynes, T. H. Geballe, G. W. Hull, Jr., and J. P. Garno, Phy. Rev. B **27**, 5188 (1983)

4 Shuichi. Ishida, Keiki. Takeda, Atsushi Okamoto, Ichiro Shibasaki, Physica. E **20**, 211

(2004)

5 S. Hikami, A. Larkin and Y. Nagaoka, Prog. Theor. Phys. 63, 707 (1980)

6 S. V. Iordanskii, Y. B. Lyanda-Geller, G. E. Pikus, JETP Lett. **60**, 206 (1994)

7 A. Kawabata, Solid State Commun. **34**, 431 (1980)

8 H. Fukuyama, K. Hoshino, J. Phys. Soc. Jpn. **50**, 2131 (1981)

9 Christopher Schierholz, Toru Matsuyama, Ulrich Merkt and Guido Meier, Phy. Rev. B **70**, 233311 (2004)

10 G. M. Minkov, A. V. Germanenko, O. E. Rut, and A. A. Sherstobitov, Phy. Rev. B **70**, 155323 (2004)

11 R. G. Mani, L. Ghenim, and J. B. Choi, Solid State Commun. **79**, 8 (1991)

12 J. Rammer, and A. Schmid, Phy. Rev. B **34**, 1352 (1986)

13 M. Yu. Reizer, Phy. Rev. B **40**, 5411 (1989)

14 F. E. Meijer, A. F. Morpurgo, T. M. Klapwijk, T. Koga, and J. Nitta, Phy. Rev. B **70**, 201307(R) (2004)

15 A. G. Mal'shukov, K. A. Chao, M. Willander, Phy. Rev. B **56**, 6436 (1997)

16 S. Cabañas, Th. Schäpers, N. Thillosen, N. Kaluza, V. A. Guzenko, and H. Hardtdegen, Phy. Rev. B **75**, 195329 (2007)